\documentclass[twocolumn,superscriptaddress,showpacs,prl]{revtex4-1}

\usepackage{amssymb,amsmath,graphicx,amsthm,wrapfig,epstopdf,esint,afterpage,marginnote,afterpage,lipsum,color,tabularx}
\usepackage{bm}
\usepackage{natbib}
\usepackage[version=4]{mhchem}
\usepackage{color,sidecap,fancyhdr}

\newcommand{\eps}{\epsilon}
\newcommand{\bra}[1]{\left(#1\right)}
\newcommand{\Bra}[1]{\left[#1\right]}

\begin{document}

\title{Mean field approach to first and second order phase transitions in ionic liquids}

\author{Sariel Bier}
\affiliation{Department of Solar Energy and Environmental Physics, Swiss Institute for Dryland Environmental and Energy Research, 
Blaustein Institutes for Desert Research (BIDR), Ben-Gurion University of the Negev, Sede Boqer Campus, 8499000 Midreshet Ben-Gurion, Israel}

\author{Nir Gavish}\affiliation{Department of Mathematics, Technion - IIT, 3200003 Haifa, Israel}

\author{Hannes Uecker} \affiliation{Institute for Mathematics, Carl von Ossietzky University of Oldenburg, P.F 2503, 26111 Oldenburg, Germany}

\author{Arik Yochelis}
\affiliation{Department of Solar Energy and Environmental Physics, Swiss Institute for Dryland Environmental and Energy Research, 
Blaustein Institutes for Desert Research (BIDR), Ben-Gurion University of the Negev, Sede Boqer Campus, 8499000 Midreshet Ben-Gurion, Israel}
\email{yochelis@bgu.ac.il}

\received{\today}

\begin{abstract}

Ionic liquids are solvent-free electrolytes, some of which possess an intriguing self-assembly property. Using a mean-field framework (based on Onsager's relations) we show that bulk nano-structures arise via type-I and II phase transitions (PT), which directly affect the electrical double layer (EDL) structure. Ginzburg-Landau equation is derived and PT are related to temperature, potential and interactions. The type-I PT occurs for geometrically dissimilar anion/cation ratio and, surprisingly, is induced by perturbations on order of thermal fluctuations. Finally, we compare the insights with the decaying charge layers within the EDL, as widely observed in experiments.

\end{abstract}

\maketitle
	
Molten salts are comprised of large and asymmetric anions and cations, with their molecular structure consisting of a charged ion attached to a hydrophilic or hydrophobic functional group. With significant charge delocalization and irregular geometries, the ions do not readily form a tightly-bound lattice and remain liquid even at room temperatures and in the absence of any solvent~\cite{Dupont2011}, hence the name ``ionic liquids" (ILs). Their tunable molecular structure enables the tailoring of ILs to a large number of applications~\cite{B006677J,SilComp:2006,Wishart2009956,B817899M,Brandt2013554,Li201483,MacFarlane2014,Armand2009621}, e.g., batteries, supercapacitors, dye-sensitized solar cells, lubricants and nanoparticle syntheses, where they are advantageous due to their high charge density, low-volatility, and high chemical, thermal and electrochemical stability.  

It is the amphiphilic-type structure, however, that gives ILs another intriguing property- the ability to \textit{self-assemble}, see~\cite{Yochelis2015} and references therein. IL molecules spontaneously form bicontinuous, hexagonal or lamellar phases, see~\cite{hayes2015structure} and references therein, in a fashion similar to the morphologies of block copolymers and liquid crystals. The bulk nano-structure effects not only the mechanical and transport properties~\cite{C5CP07090B}, but also the electrical double layer (EDL) structure and thus charge transfer properties~\cite{Endres20101724,uysal2013structural,rotenberg2015structural,fedorov2014ionic}. Obtaining insight into the emergence of nanostructure is therefore essential for the integration and control of ILs in scientific and industrial applications~\cite{B006677J,SilComp:2006,Wishart2009956,B817899M,MacFarlane2014,Armand2009621,fedorov2014ionic}.

It was recently proposed that the coupling of bulk and EDL morphologies may be portrayed as the result of competition between short-range intermolecular interactions (e.g., hydrogen bonds, solvation interactions, steric effects) and the long-range Coulombic interactions~\cite{GavishYochelis2016}, where the contribution of the former may be incorporated via the Cahn--Hilliard theory of phase separation. This nonlocal Cahn--Hilliard framework has successfully reproduced the emergence of bulk nano-morphologies that have been observed experimentally~\cite{hayes2015structure,Yochelis2015}, while providing further insights into electrokinetic phenomena in ILs, namely in the form of transient currents. However, the methodology was based on geometrically identical ions, which is clearly an abstraction of any real-world system~\cite{hayes2015structure}. Furthermore, recent experiments have implied that phase transitions at the EDL should be attributed to electrode polarization~\cite{rotenberg2015structural}. However, it is difficult to determine solely on the basis of experiments and atomistic simulations whether phase transitions in EDL structure depend only on the specific electrode polarization or whether there is an additional competing mechanism originating in the { bulk~\cite{Yochelis2014b}}. 
Isolating the separate influences of the bulk and interface properties is challenging as the coupling is inherent in the system~\cite{GavishYochelis2016}, and measurements of the EDL alone may lead to erroneous conclusions about the bulk, namely, that decaying charge layering in the EDL immediately implies an unstructured {bulk~\cite{Bazant2011}}.

In this Letter, we show that nano--patterning in ILs is a result of first and second order phase transitions. To this end, we extend the symmetric nonlocal Cahn-Hilliard mean field theory~\cite{GavishYochelis2016} to account for ion size asymmetry; the new formulation also relies on Onsager theory. Then, using weakly nonlinear analysis, we derive a Ginzburg--Landau equation and perform numerical continuation to distinguish between the types of phase transitions that account for the emergence of nano-morphology both in the bulk and in the EDL. Particularly, we show that the formation of ordered layers near the solid surfaces is induced by naturally occurring energy fluctuations, while the observed decaying spatial charge oscillations may be attributed to an isotropic bulk morphology (independent of the phase transition type), rather than an unstructured bulk.

We start by considering a fully dissociated IL of monovalent anions ($n$) and cations ($p$), where their molar concentrations range as $0\le n\le n_{\max}$ and~$0\le p\le p_{\max}$, respectively. The free energy of such an asymmetric system in the mean-field framework is generally comprised of the phase separation and Coulombic components~\cite{GavishYochelis2016}:
\begin{eqnarray}\label{eq:energy}
	&&\mathcal{E}=\int\text{d}{\bf x} \left \{k_BT \left[p\ln \frac {p}{\bar c} +n\ln \frac {n}{\bar c}\right]+\frac{\bar c\beta}{p_{\rm max}n_{\rm max}} np\, + \right. \\
\nonumber		&&\left. \frac{E_0\kappa^2\bar{c}}4 \left(\left|\frac{\nabla p}{p_{\rm max}}\right|^2+\left|\frac{\nabla n}{n_{\rm max}}\right|^2\right) + q\bar{c}(p-n)\phi-\frac12\epsilon|\nabla \phi|^2 \right\},
\end{eqnarray}
where,~$\bar{c}$ is the harmonic average density of ions,~$k_B$ is the Boltzmann constant,~$T$ is the temperature,~$\beta$ is the interaction parameter for the anion/cation mixture and has units of energy,~$E_0\kappa^2/4$ is the interfacial energy coefficient where~$E_0$ has units of energy and~$\kappa$ has units of length, $\phi$ is the electric potential,~$q$ is the elementary charge, and~$\epsilon$ is the permittivity. {The first three 
terms in~\eqref{eq:energy} correspond to the Flory--Huggins functional}, which for $\beta>\beta_c=4k_BT$ takes the form of a double well potential, thus driving phase separation~\cite{huggins1942theory,Flory_book,cahn1958free},  while {the last two terms of~\eqref{eq:energy}} are related to electrostatics,{i.e., Poissons' equation}.

The phase transitions to bulk morphology are related to spatiotemporal symmetry--breaking of a uniform bulk concentration. To derive the dynamic equations for the charge densities, we exploit the Onsager framework~\cite{onsager1932irreversible,hilliardspinodal} 
\begin{equation}\label{eq:motion_gen}
	\dfrac{\partial}{\partial t}\left(\begin{array}{c} p\\n\end{array}\right)=\frac{M}{\bar{c}}\nabla\cdot \left[pn \left(\begin{array}{rr}\gamma&-1\\-1& 1/\gamma\end{array}\right) \nabla \left(\begin{array}{c} \delta {\mathcal E}/\delta p\\ \delta {\mathcal E}/\delta n \end{array} \right)
	\right],
\end{equation}
where $M$ is the diffusion coefficient~\cite{de1980dynamics,brochard1983polymer,psaltis2011comparing} and $\gamma=p_{\max}/n_{\max}>0$ is the packing ratio of the ions affected by their geometry; $\gamma=1$ corresponds to the symmetric (1:1) case where $p_{\max}=n_{\max}=1/2$~\cite{GavishYochelis2016}.
For the sake of analysis, we introduce non-dimensional variables and parameters
$\tilde p=p/p_{\max}$, $\tilde n=n/n_{\max}$, $\bf{\tilde x}=\bf{x}/\lambda$, $\tilde t=t/\tau$, $\tilde{\phi}=q/\bra{k_BT}\phi$, $\lambda= \sqrt{\gamma \epsilon k_B T/\Bra{\bra{1+\gamma}q^2 {c}_{\max}} }$, $\tau=\lambda^2/\Bra{(1+\gamma)K_BT}$, and obtain (after omitting the tildes) the final dimensionless equations for the asymmetric ILs: 
\begin{subequations}\label{eq:rtil_model}
\begin{eqnarray}
\label{eq:rtil_model_p}
	\partial_t p&=&\nabla\cdot \left[\left(1+\frac{1-\gamma}{\gamma}p\right) \nabla p + \right.  \\
\nonumber	&&\left. p(1-p)\left(\frac{1+\gamma}{\gamma}\nabla \phi-\frac{2\chi}{1+\gamma}\nabla p-\frac{2\sigma}{1+\gamma}\nabla^3 p\right)\right],\\
\nabla^2 \phi&=&1-(1+\gamma)p,
\label{eq:rtil_model_phi}
\end{eqnarray}
\end{subequations}
with $\partial_t n=-\gamma^{-1} \partial_t p$. The parameter $\sigma=\frac{E_0}{2k_BT}\frac{\kappa^2}{\lambda^2}$ depends on the ratio of the strengths of short and long-range interactions and reflects the competition between them, and $\chi=\beta/(k_BT)$ is the Flory parameter. For consistency with traditional dimensionless analysis, we have chosen to scale space by the Debye-like scale~$\lambda$ while noting that this choice does not reflect a typical electric screening length, as for dilute electrolytes. 

\begin{figure}[tp]
	(a)\includegraphics[width=0.4\textwidth]{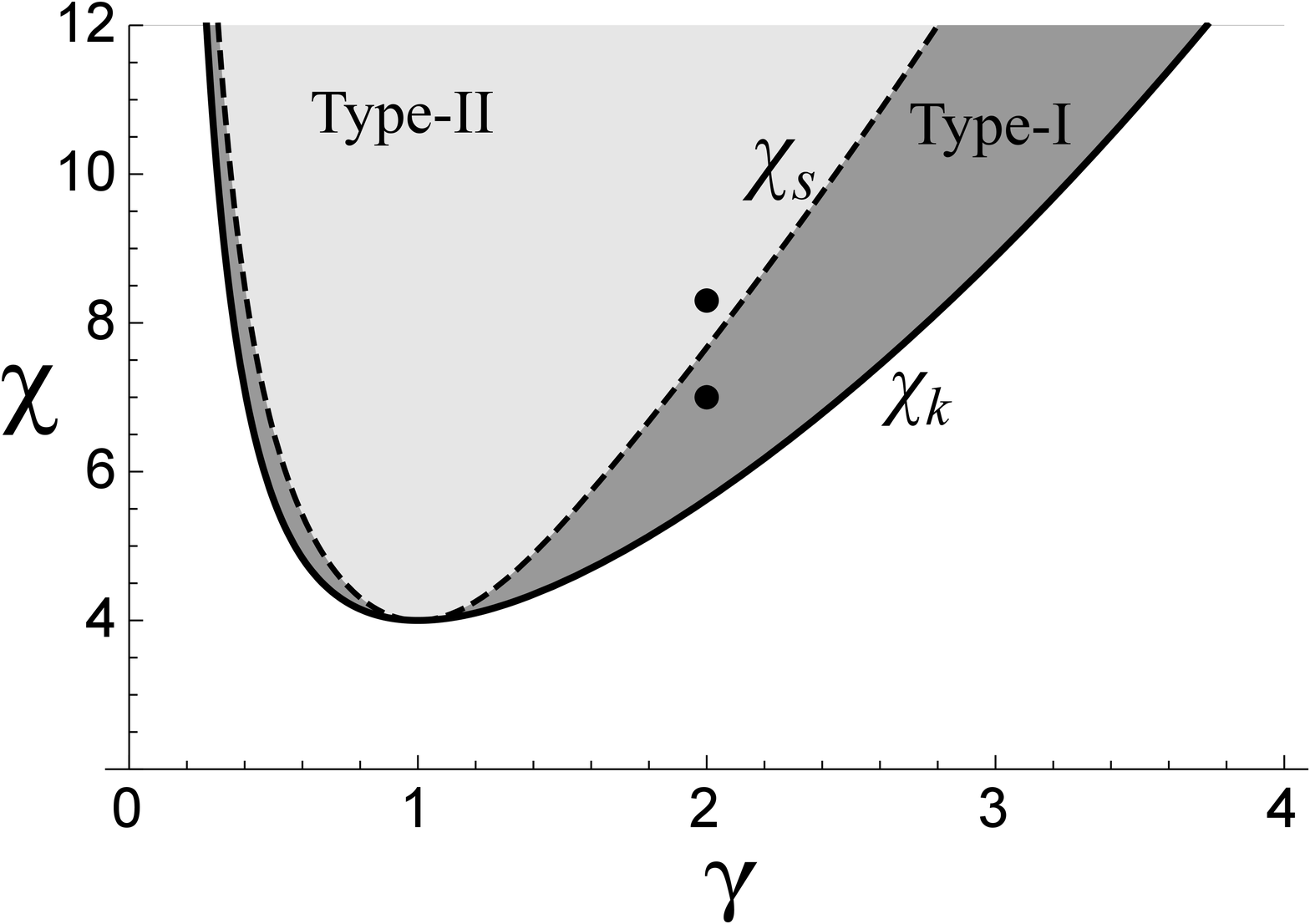}
	(b)\includegraphics[width=0.4\textwidth]{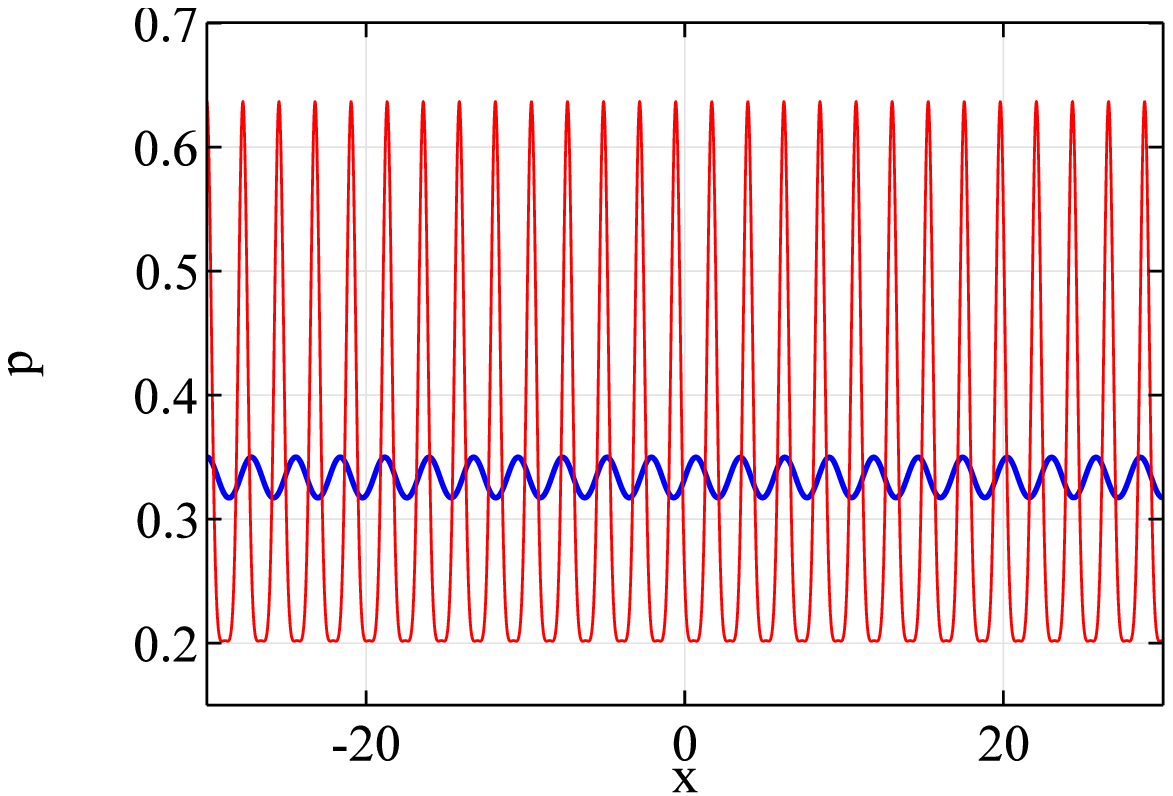}
	\caption{(color online) (a) Parameter space depicting the regions of parameters $\gamma,\chi$ in which the system undergoes type-I (subcritical) and type-II (supercritical) phase transitions {at $\sigma=\sigma_c$}. $\chi_k$ and $\chi_S$ are given by~\eqref{eq:chi_k} and~\eqref{eq:chi_s}, respectively. (b) Asymptotic charge profiles obtained by direct numerical integration of~\eqref{eq:rtil_model} for $\chi=8.3$ (light color line) and $\chi=7$ (light color line) at $\gamma=2$ and $(\sigma-\sigma_c)/\sigma_c=-0.001$, as also marked by solid circles in (a), respectively; initially $p(x,t=0)=p_0$ and $\phi(x,t=0)=0$ with effective no-flux boundary conditions where employed for all variables except $\phi(L=30)=0$. 
	} \label{fig:1}
\end{figure}

{The emergence of spatially inhomogeneous solutions for \eqref{eq:rtil_model}, see, e.g., Figure~\ref{fig:1}(b), is associated with the finite wavenumber instability about a uniform solution 
as in the symmetric ($\gamma=1$) case~\cite{GavishYochelis2016}.} For 
the sake of analysis, we first consider an infinite domain in one space dimension, using $\sigma$ as a control parameter. We then proceed to demonstrate the validity of the results on a finite domain 
$x\in(-L,L)$ as well, {using no-flux (i.e., Neumann) boundary conditions for $p$, and  fixed potential (i.e., Dirichlet) boundary conditions \textbf{$\phi(x=\pm L)=\pm V/2$}, where $V$ is the applied potential difference.}

Employing the weakly nonlinear theory (multiple time scale analysis), we expand 
\begin{equation}\label{eq:expansion}
\begin{pmatrix}
p \\ \phi
\end{pmatrix}=
\begin{pmatrix}
p_0 + \epsilon^{1/2}p_1+\epsilon p_2 +\epsilon^{3/2}p_3+... \\
\epsilon^{1/2}\phi_1 +\epsilon  \phi_2 +\epsilon^{3/2}\phi_3+...
\end{pmatrix},
\end{equation} 
where $p_0=1/(1+\gamma)$ is the uniform cation concentration, $\epsilon \ll 1$ measures the distance from the instability onset, and following~\cite{shiwa1997amplitude,gavish2016spatially} 
\[
p_1=A(\eps t,\sqrt{\eps}x)\exp(ik_cx)+\text{complex conjugate},
\] 
and $ \phi_1=p_1/(p_0 k_c^2)$. The complex amplitude $A$ varies slowly in time, and $k_c$ is the critical wavenumber at the instability onset $\eps:=\bra{\sigma-\sigma_c}/\sigma_c=0$. Both quantities are obtained via the linear stability analysis about the uniform state {$(p,\phi)=(p_0,0)$} to periodic perturbations~\cite{Cross1993851}. {Linearizing \eqref{eq:rtil_model} 
	around $(p_0,0)$, i.e., setting $p=p_0+q$ and retaining linear terms 
	in $(q,\phi)$ yields 
	\begin{align*}
		\partial_t q=&\partial_x\biggl[\frac{1{+}\gamma^2}{\gamma(1{+}\gamma)}\partial_x q
		-\frac \gamma{(1{+}\gamma)^2}\left(\frac{2\chi}{1{+}\gamma}\partial_x q{+}\frac{2\sigma}{1{+}\gamma}\partial_x^3 q\right)\biggr]\\
		&{+}\frac 1{1{+}\gamma}\partial_x^2\phi,\\
		\partial_x^2\phi=&-(1{+}\gamma)q, 
	\end{align*}
	and substituting the second into the first equation 
	yields the dispersion relation {of finite wavenumber instability type~\cite{GavishYochelis2016}}}
\begin{equation}
s=-1+\Bra{\chi\frac{2\gamma}{\bra{1+\gamma}^3}-\frac{1+\gamma^2}{\gamma \bra{1+\gamma}}}k^2-\sigma \frac{2\gamma}{\bra{1+\gamma}^3}k^4. 
\end{equation}
$s$ is the temporal growth rate of periodic perturbations associated with {non vanishing} wavenumbers $k$, i.e., $s(k)>0$ implies linear instability. 
{Varying $\sigma$, the instability onset is obtained by seeking 
$\sigma=\sigma_c$ (the phase transition point) and $k_c$ such that $s(k=k_c,\sigma=\sigma_c)=0$, $s(k \neq k_c)<0$ and  $\frac{\mathrm{d}s}{\mathrm{d}k}(k_c)=0$, yielding}
\[ \sigma_c=\frac{2\gamma^2 \chi -(1+\gamma^2)(1+\gamma)^2)^2}{8\gamma^3(1+\gamma)^3},\quad  
k^2_c=\sqrt{\frac{(1+\gamma)^3}{2\sigma_c\gamma}}.
\]
{Notably, although the model equations are gradient the coupling to Poisson equation damps the large scale mode~\cite{shiwa1997amplitude,gavish2016spatially,GavishYochelis2016}, i.e., $s(0)<0$.}  
Since $k^2_c,\gamma>0$, the instability may exist only for 
\begin{equation}\label{eq:chi_k}
\chi>\chi_k=\frac{\bra{1+\gamma^2}\bra{1+\gamma}^2}{2\gamma^2},
\end{equation}
as identified by the shaded region in Fig.~\ref{fig:1}. 
For the symmetric case ($\gamma=1$) we retain the result $\chi_k=\chi_c\equiv \beta_c/(k_B T)=4$ {from \cite{GavishYochelis2016}. }

Slightly above the instability onset, $\eps \ll 1$, the dispersion relation admits a finite band of wavenumbers around $k_c$ for which $s>0$. However, direct numerical integration shows that at $\epsilon=-0.001$ and $\gamma=2$, we find two distinct cases: for $\chi=8.3$ the solutions correspond to small amplitude periodic solutions (see dark line in Fig.~\ref{fig:1}(b)), while for $\chi=7$ the solutions are highly nonlinear (as evidenced by the spike-like shape of the light line in~\ref{fig:1}(b)) and of large amplitude (relative to $\epsilon$). This behavior indicates two distinct distinct phase transitions mechanisms, {which can be analyzed using the amplitude equation for $A$}.  
Thus we substitute~\eqref{eq:expansion} into~\eqref{eq:rtil_model}, and after imposing the solvability condition at order $\epsilon^{3/2}$ obtain 
\begin{equation}\label{eq:amplitude_equation}
\frac{\partial A}{\partial t}=\frac{\sigma_c-\sigma}{\sigma_c} A +\alpha |A|^2A+\frac{4}{k_c^2}\frac{\partial^2 A}{\partial x^2}
\end{equation}
where
$\alpha=(1+\gamma)^2/\gamma+k_c^2(2\sigma_c k_c^2 -2\chi)/(1+\gamma)+ \frac{2(\gamma^3-1)k_c^2}{9\gamma}\left[\frac{\gamma^2-1}{2\gamma}-\left(\frac{1-\gamma}{\gamma}+\frac{2\chi(1-\gamma)}{(1+\gamma)^2}\right)k_c^2+7\frac{2\sigma_c(1-\gamma)}{(1+\gamma)^2}k_c^4\right].$
The Ginzburg--Landau partial differential amplitude equation 
\eqref{eq:amplitude_equation} can be used to approximate large scale modulations of the basic pattern 
$\cos(k_c x)$ near the (phase transition) onset~\cite{shiwa1997amplitude,gavish2016spatially}
but here for simplicity we assume $A$ to be independent of space. Then  
\begin{subequations}\label{eq:apprx}
\begin{eqnarray}
\label{eq:apprx_p}
p\simeq \frac{1}{1+\gamma}+2\sqrt{\frac{\sigma-\sigma_c}{\sigma_c \alpha}}\cos{k_cx},\\
\phi\simeq 2\frac{1}{p_0k_c^2}\sqrt{\frac{\sigma-\sigma_c}{\sigma_c \alpha}}\cos{k_cx}.
\label{eq:apprx_phi}
\end{eqnarray}
\end{subequations}

\begin{figure}[h!]
	(a)\includegraphics[width=0.4\textwidth]{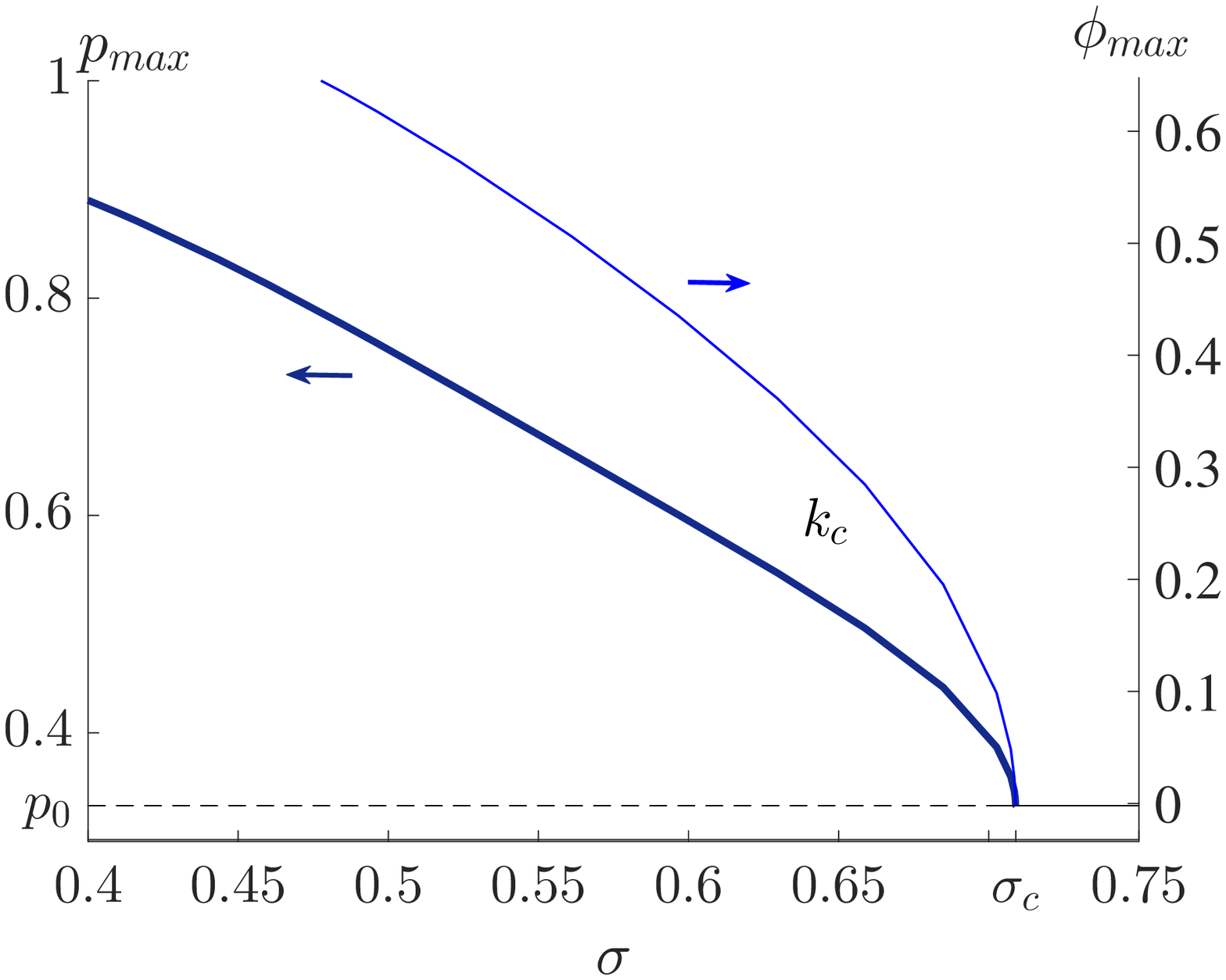}
	(b)\includegraphics[width=0.4\textwidth]{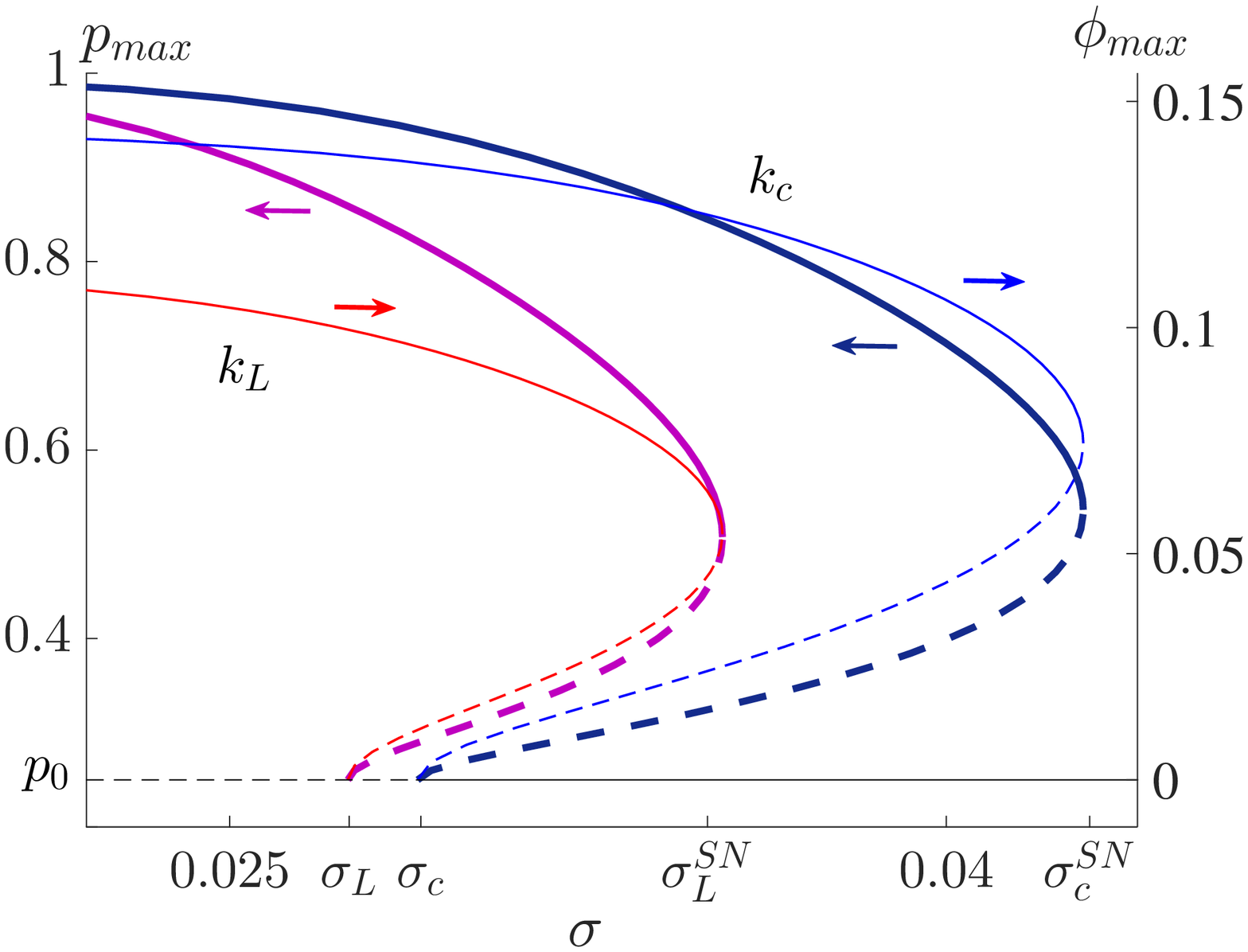}
	\caption{(color online) Bifurcation diagrams for (a) $\sigma<\sigma_c$, $\alpha<0$ and (b) $\sigma>\sigma_c$, $\alpha>0$, showing the maximal values,{~$p_{\max}=\max|p|$ (thick curves {and left arrows}) and~$\phi_{\max}=\max|\phi|$ (thin curves {and right arrows}),} of the bifurcating periodic solutions as a function of $\sigma$. The (a) super-- and (b) sub--critical branches denote the type-II and type-I phase transitions, respectively, and have been obtained using numerical continuation via the {\tt pde2path} package~\cite{pde2path}; solid (dashed) curves mark stable (unstable) solutions along the branches. The $k_c$ branch in (b) corresponds to a single-period domain size  ($2L=2\pi/k_c$), with $k_c\simeq 4.382,\sigma_c\simeq 0.029,\sigma_c^{SN}\simeq 0.043$, while the $k_L$ branch 
with $k_L\simeq 5,\sigma_L\simeq 0.027$ and $\sigma_L^{SN}\simeq 0.035$ is {one of the branches} obtained for a large domain ($2L=100$).} \label{fig:2}
\end{figure}

{From the right hand sides of~\eqref{eq:apprx}} it is evident that the signs of $\sigma-\sigma_c(\gamma,\chi)$ and of $\alpha(\gamma,\chi)$ differentiate between two types of periodic solutions: (\textit{i}) The super--critical (phase transition of type-II) bifurcation when $\sigma<\sigma_c$ and $\alpha<0$, with the amplitude scaled as $\sqrt{|\sigma-\sigma_c|}$ (see Fig.~\ref{fig:2}(a)), and (\textit{ii}) The sub--critical (phase transition of type-I) bifurcation when $\sigma>\sigma_c$ and $\alpha>0$, with the amplitude exhibiting a discontinuous jump to amplitudes $\mathcal{O}(1)$ for $\sigma<\sigma_c$. Typical solutions of both cases for $\sigma \lesssim \sigma_c$ obtained via direct numerical integration are depicted in Fig.~\ref{fig:1}(b), while the transition between the two types is obtained by solving for $\alpha(\chi,\gamma)=0$, {which yields}  
\begin{eqnarray}\label{eq:chi_s}
\nonumber \chi_s &=& \Bra{\gamma  (\gamma  (\gamma  (\gamma  (\gamma  (13 \gamma -9)+18)-26)+27)-9)+22}\\
&&\times \frac{(\gamma +1)^2}{18 \gamma ^2 \left(\gamma(\gamma -1) \left(\gamma ^2+1\right)+2\right)},
\end{eqnarray}
as depicted by the dashed line in Fig.~\ref{fig:1}(a). 

The branches of bifurcating solutions beyond the weakly nonlinear limit are computed in both cases with numerical continuation method~\cite{pde2path}, as depicted in Fig.~\ref{fig:2}. While in the super-critical case (type II) the bifurcating solutions are stable (Fig.~\ref{fig:2}(a)), in the sub-critical case (type I) these solutions are linearly unstable until reaching a fold, $\sigma_c^{SN} $, whereon they change direction and grow toward the instability onset of the homogeneous state. However, as in the Ohta-Kawasaki case~\cite{gavish2016spatially}, the large amplitude periodic solutions belonging to a branch portion after the fold ($\sigma<\sigma_{SN}$, see figure) are stable on $2L=\lambda_c=2\pi/k_c$ (exhibiting hysteresis) but become Eckhaus unstable on large domains; due to applications our interest here is in domains that are larger than the wavelength of the bifurcating states, i.e., $L \gg  \pi/k_c$. Consequently, the hysteresis region $\sigma_c<\sigma<\sigma^{SN}_c$, between the 
 $k_c$ solutions and the uniform state is destroyed on large domains. Nevertheless, direct numerical integration shows that there are additional periodic solutions that emerge and form hysteresis for $\sigma_c<\sigma<\sigma_L^{SN}$ (Fig.~\ref{fig:1}(b)). These solutions 
{belong} to one of the secondary branches of periodic solutions {(denoted by $k_L$)} that emerge below the onset, i.e., at $\sigma<\sigma_c$ as shown by thick line in Fig.~\ref{fig:2}(b).
\begin{figure}[tp]
	\includegraphics[width=0.4\textwidth]{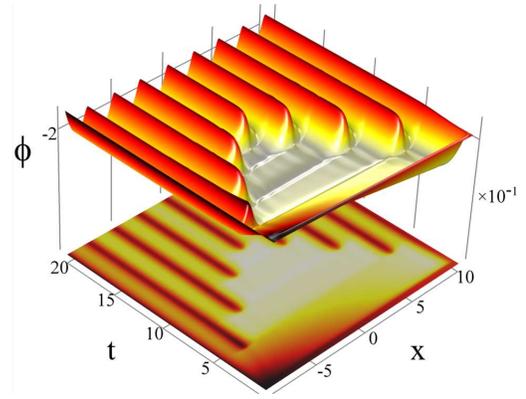}
	\caption{Space--time plot showing the formation of large amplitude solutions from a weak perturbation at the boundaries, $\phi(x=\pm L)=\pm 0.01$. The top panel shows the profile of $\phi$ as a surface while the bottom panel as a contour with initially $p(x)=p_0$ and $\phi=0$. Parameters: $\gamma=3,\chi=10$, $L=10$, $(\sigma-\sigma_c)/\sigma_c=0.1$.} \label{fig:3}
\end{figure}

From a 
mathematical point of view, bistability of the uniform and periodic state persists on finite domains. However, in the physical context, perturbations are related to energy fluctuations which operate at the scale of $k_BT$ or $\phi\sim1$ in dimensionless units. {The size of fluctuations should be 
compared to the amplitude of the unstable parts of the 
subcritical branches. Setting $\phi\approx 1$ in~\eqref{eq:apprx_phi} yields
\begin{equation}\label{eq:subV1}
\sigma-\sigma_c>\left|\frac{\alpha \sigma_c p_0^2 k_c^4}{4}\right| 
\sim  \frac{32}{\chi-4}
+\mathcal{O}\bra{\gamma-1}.
\end{equation}
The numerical continuation in Fig.~\ref{fig:2}(b) shows that 
in the parameter regimes considered here the folds are too close 
to $\sigma_c$ for \eqref{eq:subV1} to be fulfilled. Indeed, even a weak perturbation $\phi=V$ with $|V|\ll 1$ at a boundary  will result in a large--amplitude pattern after a temporal transient, as shown in Fig.~\ref{fig:3}. In other words, in a physicochemical system with the required temperature and anion/cation asymmetry for first order phase transitions, energy fluctuations will drive the formation of nano-structures for $\sigma_c<\sigma<\sigma_L^{SN}$, even though the homogeneous state is mathematically stable in this domain. } 

Finally, the large amplitude periodic patterns in 1D take on a much richer form in 3D, such as laminar, bicontinuous, and sponge-like morphologies~\cite{Cross1993851,choksi2005diblock,nyrkova1994microdomain}. Thus, the isotropic neutral bulk region is averaged upon projection on the parallel axis in between the electrodes to $(p,\phi)=(p_0,0)$, which is also consistent with the electroneutrality. Near the electrodes the isotropy is broken, and the periodic states align with the surface~\cite{GavishYochelis2016}, with the alignment more pronounced closer to the electrode. Consequently, experimental measurements such as atomic force microscopy (AFM)~\cite{Atkin201444}, atomic force apparatus~\cite{Perkin20125052} and X-ray reflectivity~\cite{uysal2013structural}, which are essentially 1D methods, observe only charge layering. We emphasize that while charge layering is an aspect phase transition of type-I in the bulk and loss of isotropy near the electrodes, it does not
  automatically imply bulk nano-structure. The distinction between structured and unstructured bulk may be further advanced, for example, using time dependent measurement which shows different behavior in each case~\cite{GavishYochelis2016}. 
In higher spatial dimensions the 1D periodic bulk structure in takes the universal forms of labyrinthine and hexagonal patterns. The former may be expected in the region of type-II phase transitions for nearly--symmetric cases ($\gamma \approx 1$) and emerges via a zigzag instability~\cite{GavishYochelis2016}, while the latter emerges via a sub-critical bifurcation~\cite{GoNep:06}. In both cases, similarly to the first order phase transition region, perturbations on the order of $\phi=1$ will cause the emergence of periodic morphology that will average to electroneutrality in the bulk.

To conclude, we have demonstrated the impact of cation/anion size asymmetry on ionic liquid self-assembly in both the bulk and the electrical double layer regions. {Importantly, the framework is consistent with the second law of thermodynamics through the nonlocal Cahn-Hilliard approach that is based on Onsager's relations~\cite{GavishYochelis2016}. The nano-morphology is shown to arise first in the bulk via phase transitions of types I and II under certain conditions, following the derived universal Ginzburg-Landau equation~\eqref{eq:amplitude_equation}}. In both types, near the transition threshold, $\sigma_c$, the bulk is sensitive to energy fluctuations of about $k_BT$ or respectively $25mV$, with the resulting formation of large amplitude nano-morphologies even in the type I case. Since these phase transitions are of universal nature~\cite{Cross1993851,GoNep:06}, in higher space dimensions the morphologies that form are of isotropic nature (e.g., labyrinths and hexagons) and thus, upon averaging about the parallel plane to the electrodes, correspond to a uniform density with a vanishing electrical field, i.e., electroneutrality in 1D context. The second effect {of an isotropic structured bulk is its} subtle impact on the charge layering within the EDL region~\cite{Yochelis2015,hayes2015structure}. Since the bulk isotropy breaks down near the solid surface, the patterns tend to align with electrode orientation. Upon averaging about the plane normal to the electrode one naturally observes charge layers near the electrode that gradually lose their orientation toward the bulk. Consequently, spatially decaying oscillations are observed in the direction normal to the electrode, yet they should not be misinterpreted as evidence for absent bulk structure. As many applications~\cite{hayes2015structure,fedorov2014ionic}, e.g., energy- and lubrication-related, exploit and depend on mass transport and charge transfer properties, the framework developed here offers a new perspective regarding the interpretation of empirical observations, as well as opportunities for the enhancement of device efficiency~\cite{B006677J,SilComp:2006,Wishart2009956,B817899M,Brandt2013554,Li201483,MacFarlane2014,Armand2009621} in terms of conductivity, structural integrity (rheology), and electrochemical reactions.

\bibliographystyle{apsrev4-1}
\bibliography{IL2}
\end{document}